\documentclass{aa}
\usepackage{graphicx}
\usepackage{hyperref}
\usepackage{amsmath}
\usepackage[varg]{txfonts}
\usepackage{color}
\usepackage[english]{babel}

\newcommand{\cm}{cm$^{-1}$}

\def\a0{{$a_{\rm 0}$}}

\titlerunning{ExoCross}
\authorrunning{Yurchenko et al.}

\newcommand{\tnu}{\tilde{\nu}}
\newcommand{\tE}{\tilde{E}}
\newcommand{\dnu}{\ensuremath{\Delta\tilde{\nu}}}
\newcommand{\rmd}{{\rm d\,}}

\newcommand{\xcross}{\textsc{ExoCross}}

\newcommand{\Humlicek}{Huml\'{i}\v{c}ek}

\graphicspath{ {eps/} }

\begin{document}

\title{{\sc ExoCross}: a general program  for generating spectra from molecular line lists}

\author{Sergei N. Yurchenko \thanks{The corresponding author: Sergei N. Yurchenko; E-mail: s.yurchenko@ucl.ac.uk} \inst{1}
 \and Ahmed~F. Al-Refaie\inst{1} \and Jonathan Tennyson \inst{1}}

\institute{%
Department of Physics and Astronomy, University College London, London WC1E 6BT, United Kingdom}

\date{\today}

\abstract{
{\sc ExoCross} is a Fortran code for generating spectra (emission,
absorption) and thermodynamic properties (partition function, specific heat etc.)
from molecular line lists. Input is taken in several formats, including
ExoMol and HITRAN formats. {\sc ExoCross} is efficiently parallelized
showing also a high degree of vectorization. It can work with several line
profiles such as Doppler, Lorentzian and Voigt and support several broadening
schemes. Voigt profiles are handled by several methods allowing fast and
accurate simulations. Two of these methods are new. {\sc ExoCross} is also capable of working with the recently proposed method of super-lines. It supports calculations of lifetimes, cooling
functions, specific heats and other properties. {\sc ExoCross} can be used to convert between different formats, such as HITRAN,
 ExoMol and Phoenix. It is capable of simulating non-LTE spectra using a simple
two-temperature approach. Different electronic, vibronic or vibrational bands
can be simulated separately using an efficient filtering scheme based on the
quantum numbers.
}

\maketitle

\titlerunning{ExoCross}
\authorrunning{S.N. Yurchenko et. al.}



\section{Introduction}
\label{sec:intro}

We present a Fortran 2003 program \xcross\ to compute spectra as well
as spectral properties of molecules using line lists. \xcross\ is
specifically developed to work with huge molecular line lists such as
those generated as part of our ExoMol project \citep{12TeYuxx.db} or
similar endeavours \citep{TheoReTS}.  \xcross\ takes such line lists
as input and returns pressure- and temperature-dependent cross
sections as well a variety of other derived molecular properties which
depend on the underlying spectroscopic data.  These include
state-dependent lifetimes, temperature-dependent cooling functions,
and thermodynamic properties such as partition functions and specific
heats.

The main challenge when working with hot line lists for polyatomic
molecules is their extremely large sizes.  Thus, for example, there
are several line lists generated as part of the ExoMol project
containing in excess of 10 billion transitions
\citep{jt564,jt592,jt641,jt698,jt701,15PaYuTea,15AlYaTe.H2CO,15AlOvPo.H2O2,jt698}.
The size of these datasets makes them impractical for direct use in
line-by-line applications.  We note that simply ignoring the billions
of often very weak lines does not give reliable results
\citep{jt572,jt698}.  While there are a number of approaches to this
problem such as the use of $k$-coefficients (see, for example,
\citet{09ShFoLi,14AmBaTr.broad,HELIOS,17Min}), the most practical
approach which does not involve making significant approximations is
to produce cross sections for a set of predefined conditions. These
cross sections are then easier to handle in, for example, radiative
transfer codes than the original line lists as they can be stored on
far fewer grid points than there are lines.  However handling these
large line lists requires care and, in particular, the generation of
cross sections on an appropriate temperature-, pressure- and
frequency/wavelength-dependent grid is data intensive and can become
computationally highly demanding. \xcross\ provides a computational
solution to this problem; it has been extensively optimised to process
huge datasets, including the introduction of an efficient algorithm
for generation large numbers of Voigt profiles which is discussed
below. \xcross\ is optimized to provide high throughput via efficient
parallelization and vectorization. This is especially important when
working with line lists containing tens of billions lines.  At
different stages of development \xcross\ was used to generate spectra
by
\citet{jt635,jt641,jt644,jt686,jt693,jt698,jt701,jt703,jt711} and \citet{jt712}.

\xcross\ is designed to generate molecular cross sections
(absorption or emission) on a grid for a set of temperatures and  pressures using
different line profiles (e.g. Doppler, Voigt etc) under the local thermodynamic equilibrium (LTE) as well as non-LTE \citep{jt722}. Other useful
functionality include computing lifetimes \citep{jt624}, stick spectra, partition functions, cooling functions, and specific heats.
The HITRAN molecular spectroscopic database \citep{jt691} is a widely-used compilation aimed at radiative
transport studies of the Earth's atmosphere.
\xcross\ is capable of working with HITRAN line lists (\texttt{.par}) as well as super-lines \citep{TheoReTS,jt698}.  It can be easily extended to accept other formats.

As part of this implementation, we have developed two new algorithms to perform  convolution integrals needed for the Voigt line profile. The first algorithm is based on the Gauss-Hermite quadratures and is developed specifically to guarantee conservation of the Voigt line area. The second algorithm is based on exploiting the similarity of the Voigt profile at large distances from the line centre to compute the opacities quickly.

There are a number of other similar
 programs available which are  designed to work with line lists. These include the HITRAN interface \textsc{HAPI} \citep{HAPI},
\textsc{SpectraPlot.com} \citep{SpectraPlot.com} and \textsc{SPECTRA} \citep{jt128}.
However, all of these programs would struggle to handle the huge line lists required for models
of atmospheres at elevated temperatures. \xcross\ is designed to be flexible; it takes input
in both ExoMol \citep{jt548,jt631} and HITRAN \citep{jt350} formats. Data can be returned in a variety
of formats: ExoMol, HITRAN and Phoenix \citep{PHOENIX}, where
Phoenix is a full non-LTE atmospheric transfer code accounting for depth-dependent abundances (cloud formation, element diffusion, etc.) using the line-by-line approach. Thus as a subsidiary function the code
can be used to interconvert between ExoMol and HITRAN formats.


The paper is organised as follows. The main functionality of \xcross\ is presented in Section~\ref{s:main}. The line profile implemented in \xcross\ are discussed in Section~\ref{s:profile}. Section~\ref{s:protocal} presents \xcross\ calculation steps. The data format are described in Section~\ref{s:format}. Section~\ref{s:conclusion} offers some conclusions. The \xcross\ manual provided as part of the supplementary data as well as GitHub and CCPForge repositories
gives full working details of the program so the description below
is restricted to outlines and examples.

\section{Main functionality}
\label{s:main}

\subsection{Intensities and partition function}

An absorption line intensity $I_{\rm fi}$ (cm/molecule), also known as absorption coefficient, is given by
\begin{equation}
\label{e:int}
 I({\rm f} \gets {\rm i}) = \frac{g_f^{\rm tot} A_{\rm fi}}{8 \pi c \tnu_{\rm fi}^2}  \frac{e^{-c_2 \tilde{E}_i/T } \left( 1-e^{-c_2\tilde{\nu}_{\rm fi}/T} \right)}{Q(T)},
\end{equation}
where $A_{\rm fi}$ is the Einstein-A coefficient ($s^{-1}$), $\tilde{\nu}_{\rm fi}$ is the transition wavenumber (\cm), $Q(T)$ is the partition function defined as as sum over states
\begin{equation}
\label{e:pf}
  Q(T) =\sum_{n}  g_n^{\rm tot} e^{-c_2\tilde{E}_n/T},
\end{equation}
$g_n^{\rm tot}$ is the total degeneracy given by
$$
g_n^{\rm tot} = g^{\rm ns}_n (2 J_n+1),
$$
$J_n$ is the corresponding total angular momentum, $g^{\rm ns}_n$ is the nuclear-spin statistical weight factor, $c_2= hc / k_B$
is the second radiation constant (cm K), $\tilde{E}_i = E_i/h c $ is the energy term value (\cm), and $T$ is the temperature in K.

The emissivity (erg/molecule/sterradian) is  given by:
\begin{equation}
\label{e:emiss}
 \epsilon({\rm i} \to  {\rm f}) = \frac{g_{\rm i}^{\rm tot} A_{\rm fi} \tnu_{\rm fi}}{4 \pi }  \frac{e^{-c_2 \tilde{E}_{\rm i}/T }}{Q(T)}.
\end{equation}


Note that the isotopic abundance is not included in the definition of the line intensities (absorption or emission) in Eqs.~\eqref{e:int} and \eqref{e:emiss}. This is different from the HITRAN convention, where the absorption coefficients of an isotopologue contain the corresponding natural (terrestrial) isotopic abundances, see \\ \url{https://www.cfa.harvard.edu/hitran/molecules.html}.  For such applications where the isotopic abundance is required, the intensities in Eqs.~\eqref{e:int} and \eqref{e:emiss} can be scaled by an abundance factor  specified in the input.

\subsection{Radiative lifetime}

\begin{figure}
\begin{center}
\includegraphics[width=.94\columnwidth]{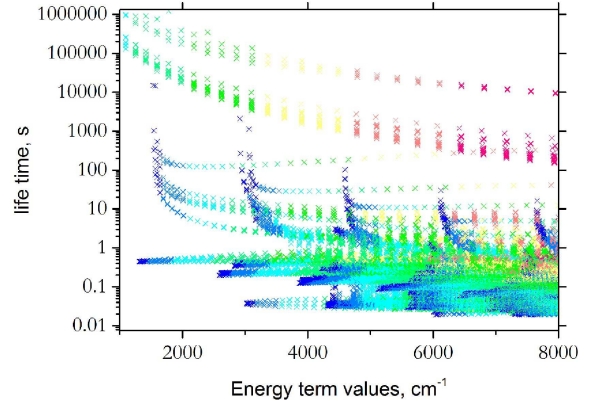}
\caption{Lifetimes of CH$_4$ computed using the 10to10 line list \citep{jt564}. The colors range from dark blue ($J=0$) to red ($J=45$). See \citet{jt624}
for a full discussion of methane lifetimes.}
\label{f:life}
\end{center}
\end{figure}

The radiative lifetime (s) can be computed as \citep{jt624}
\begin{equation}
\tau_{\rm i} =  \frac{1}{\sum_{\rm f} A_{\rm fi}}~.
\end{equation}
See an example of the lifetimes in Fig.~\ref{f:life} computed from the 10to10 line list for CH$_4$. Examples of ExoMol lifetimes and cooling functions can be found in \citet{jt624,16MeYuTe} and \citet{17MiAlZo}.

\subsection{Cooling function}

The emissivity (erg$/$s~sr~molecule) can be used to produce the cooling function
$W(T)$  as  the total
energy emitted by a molecule \citep{jt181}
\begin{equation}
W(T) =   \frac{1}{4 \pi Q(T)} \sum_{\rm f,i} A_{\rm fi} \, h c  \,\tnu_{\rm fi}  g_{\rm i} \exp\left(\frac{-c_2 \tilde{E}_{\rm i}}{T}\right).
\end{equation}


\subsection{Stick spectra}

\begin{figure}
\begin{center}
\includegraphics[width=.94\columnwidth]{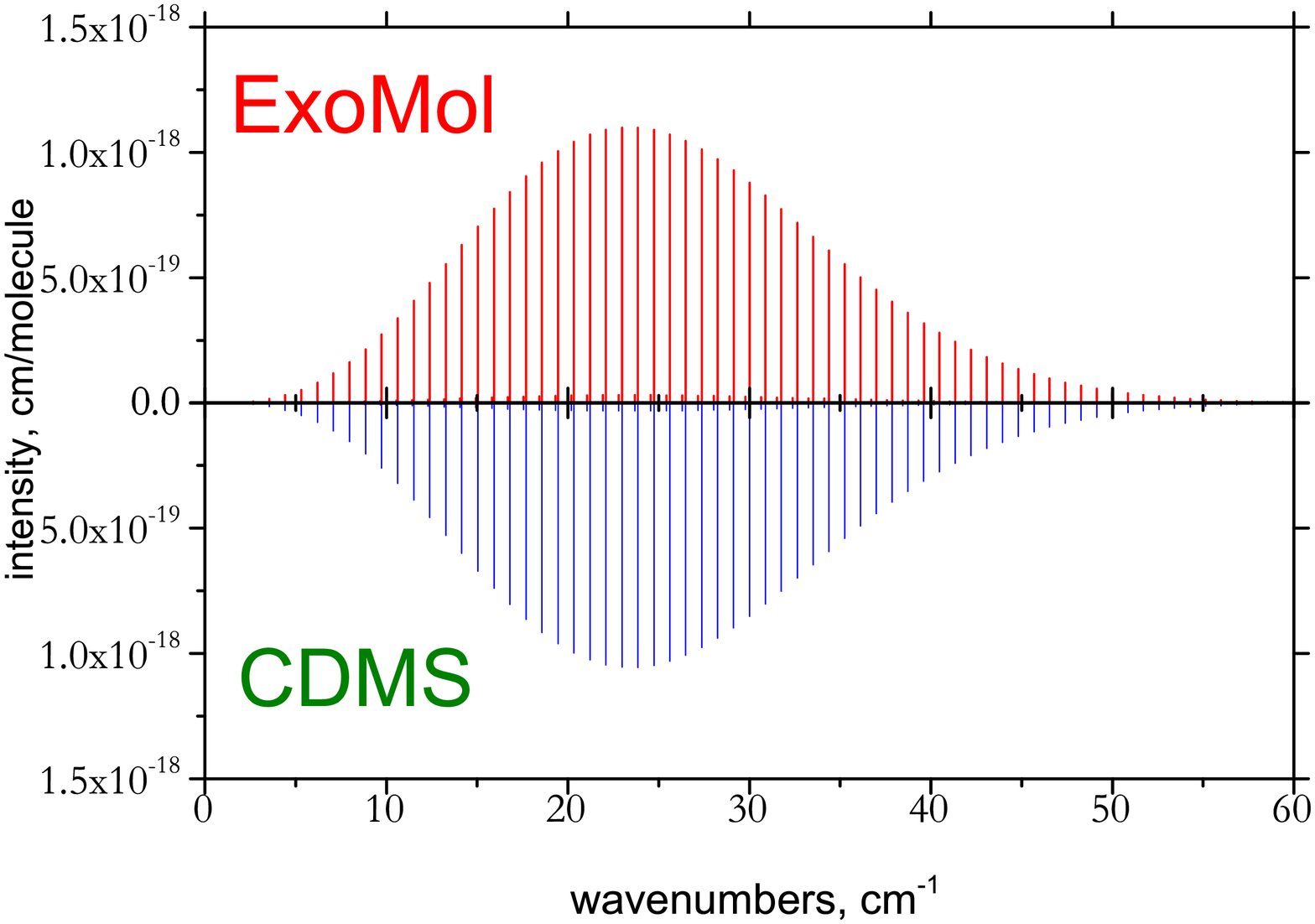}
\caption{Stick spectra (cm/molecule) of CaO \citep{16YuBlAs} compared the
 CDMS \citep{CDMS} rotational band at $T$=298~K. }
\label{f:CaO}
\end{center}
\end{figure}

A stick spectrum is a list of frequencies and line intensities,
accompanied by the full description (quantum numbers) of the upper and lower states.
When plotted, each line is represented by a `stick' with the intensity given by its height, see Table~\ref{t:stick} where an extract from an output file containing an absorption stick spectrum of KCl \citep{14BaChGo.NaCl} is shown. A stick spectra of CaO is shown in Fig.~\ref{f:CaO}.

\begin{table*}
\caption{Extract from a stick spectrum output generated using the KCl line list of \protect\citet{14BaChGo.NaCl}.  }
\label{t:stick}
\tt
\footnotesize
 \begin{tabular}{rrrrcrrrcr}
 \hline
$\tnu_{\rm fi}$ & $I_{\rm fi}$ (cm/molecule)	& $J'$ &$\tE'$& & $J''$ & $\tE''$& $\varv'$ && $\varv''$\\
\hline
   2.50210000E-01 & 3.27956194E-26  &   1   &   1096.1338  & <-   &  0 &     1095.8836   &       4  & <-  &   4 \\
   2.51764000E-01 & 1.20091306E-25  &   1   &    825.6787  & <-   &  0 &      825.4269   &       3  & <-  &   3 \\
   2.53325000E-01 & 4.44675257E-25  &   1   &    552.8938  & <-   &  0 &      552.6404   &       2  & <-  &   2 \\
   2.54891000E-01 & 1.66533127E-24  &   1   &    277.7596  & <-   &  0 &      277.5047   &       1  & <-  &   1 \\
   2.56466000E-01 & 6.30771280E-24  &   1   &      0.2565  & <-   &  0 &        0.0000   &       0  & <-  &   0 \\
   4.94245000E-01 & 2.01890019E-26  &   2   &   1630.6257  & <-   &  1 &     1630.1315   &       6  & <-  &   6 \\
   4.97324000E-01 & 7.23121652E-26  &   2   &   1364.7757  & <-   &  1 &     1364.2784   &       5  & <-  &   5 \\
   5.00417000E-01 & 2.61886416E-25  &   2   &   1096.6342  & <-   &  1 &     1096.1338   &       4  & <-  &   4 \\
   5.03524000E-01 & 9.58980316E-25  &   2   &    826.1822  & <-   &  1 &      825.6787   &       3  & <-  &   3 \\
   5.06644000E-01 & 3.55093592E-24  &   2   &    553.4004  & <-   &  1 &      552.8938   &       2  & <-  &   2 \\
   5.09779000E-01 & 1.32976733E-23  &   2   &    278.2693  & <-   &  1 &      277.7596   &       1  & <-  &   1 \\
   5.12927000E-01 & 5.03675191E-23  &   2   &      0.7694  & <-   &  1 &        0.2565   &       0  & <-  &   0 \\
   \hline
\hline

\end{tabular}
\end{table*}

\subsection{Cross sections}

A cross section $\sigma_{\rm fi}(\tilde\nu)$ from a  single line $f\gets i$ is related to the corresponding integrated absorption coefficient $I_{\rm fi}$  as
\begin{equation}\label{e:integr}
I_{\rm fi} = \int_{-\infty}^{\infty} \sigma_{\rm fi}(\tilde\nu) \;  \rmd \nu,
\end{equation}
where $\tilde\nu$  is a transitional wavenumber. By introducing a line profile $   f_{\tnu_{\rm fi}}(\tnu)$ the cross section  (cm$/($molecule cm$^{-1})$) can be defined as
\begin{equation}\label{e:cross}
\sigma_{\rm fi}(\tilde\nu) =  I_{\rm fi} f_{\tnu_{\rm fi}}(\tnu),
\end{equation}
where  $f_{\tnu_{\rm fi}}(\tnu)$ is an integrable function with the  area normalized to unity:
\begin{equation}\label{e:integr2}
\int_{-\infty}^{\infty} f_{\tnu_{\rm fi}} (\tnu) \;  \rmd \tnu = 1.
\end{equation}

\begin{figure}
\begin{center}
\includegraphics[width=.94\columnwidth]{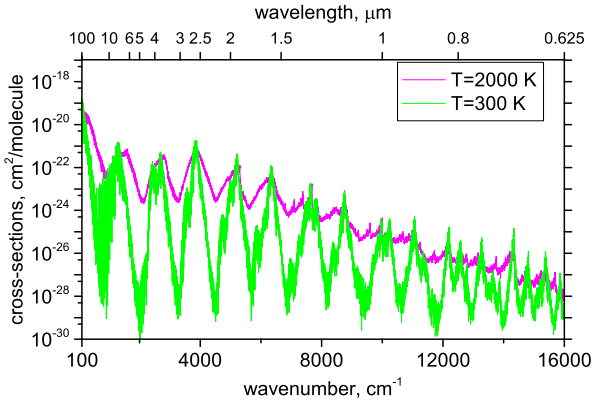}
\caption{Absorption spectrum of H$_2$S at $T$ =300~K and 2000~K simulated
using the ExoMol line list AYT2 \citep{16AzTeYu}.}
\label{f:H2S}
\end{center}
\end{figure}

Figure~\ref{f:H2S} shows an example of cross sections of H$_2$S at  $T=300$~K and $2000$~K using the ExoMol line list of \citet{16AzTeYu}.

\subsection{Grids}

By default \xcross\ uses an equidistant grid, defined by the wavenumber  of wavelength range $[\tnu_A,\tnu_B]$ and the number of the grid points $N_{\rm points}$. The latter includes both the first and last bounds. The grid bin size is defined by
\begin{equation}\label{e:Ngrid}
  \Delta \tnu = \frac{\tnu_{B}-\tnu_{A}}{(N_{\rm points}-1)}.
\end{equation}
The number of intervals is then $N_{\rm points}-1$. Usually the number of points is an odd number in order to make $\Delta \tnu $ a `round' value.

Non-equidistant wavenumbers grids can be generated either as grids of constant resolving power $R = \tnu / \Delta\tnu$  or equidistant wavelength grids.

\subsection{Partition function and specific heat}

The partition function $Q(T)$ is given by Eq.~\eqref{e:pf}. The
evaluation of $Q(T)$ requires the energy term values $\tE_i$ and
degeneracies $g_{\rm tot}$, which are usually included in molecular
line lists. As part of the intensity calculations, the partition
function must be either evaluated using these quantities, or directly
provided as part of the input. These values can be, e.g., taken from
the \texttt{.pf} files provided as part of the ExoMol database
\citep{jt631} or as part of the TIPS program provided by HITRAN
\citep{jt692}.  The direct input option is recommended as often the ExoMol or
HITRAN partition functions are more accurate as they contain
additional, higher energy contributions which make an important
contribution, particularly at elevated temperatures.

\begin{figure}
\begin{center}
\includegraphics[angle=0, width=0.94\columnwidth]{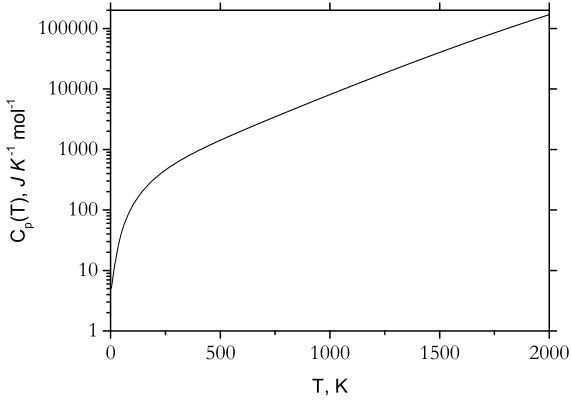}
\caption{Specific heat $C_p(T)$ of CH$_4$ computed using the 10to10 line list  \protect\citet{jt564}. \label{f:Cp}}
\end{center}
\end{figure}

The molar specific heat is given by ($J K^{-1}$ mol$^{-1}$)
\begin{equation}\label{e:Cp}
  C_p(T) = R \left[\frac{Q''}{Q}-\left( \frac{Q'}{Q}\right)  \right],
\end{equation}
where  $R$ is the gas constant and the 1st and 2nd moments $Q'$ and $Q''$  are
\begin{eqnarray*}
  Q' &=& T\frac{d Q}{d T}, \\
  Q'' &=& T^2\frac{d^2 Q}{dT^2}.
\end{eqnarray*}
These latter moments can be also requested from \xcross. An example of $C_p(T)$ of CH$_4$ generated using the 10to10 line list is shown in Fig.~\ref{f:Cp}.

\begin{figure}
\begin{center}
\includegraphics[angle=0, width=0.94\columnwidth]{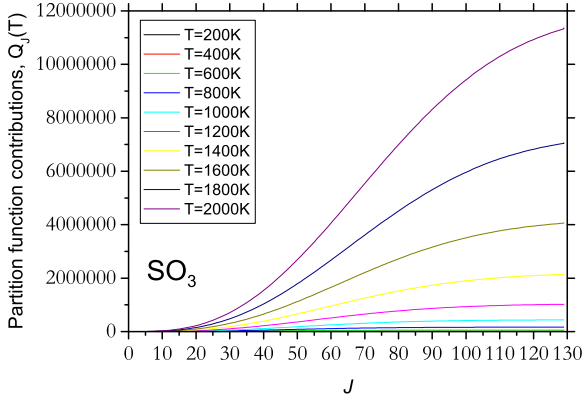}
\caption{Contributions $Q_J(T)$ to the partition function of SO$_3$ using the line list UYT2 of \protect\citet{jt641}. \label{f:SO3:Q}}
\end{center}
\end{figure}

It is often instructive to plot individual contributions to the partition function from different $J$ states defined as
\begin{equation}
\label{e:pf:J}
  Q_J(T) =\sum_{n (J)}  g_{n}^{\rm ns}  (2J_{n}+1) e^{-c_2\tilde{E}_n^{J}/T}.
\end{equation}
This is useful to assess the convergence of the line list with respect to $J$ and thus to estimate $T_{\rm max}$ the line list is applicable to. Figure~\ref{f:SO3:Q} shows the such individual $Q_J(T)$ contributions for the UYT2 line list for SO$_3$ \citep{jt641}.

\subsection{Intensity thresholds}

An intensity threshold can be used to speed up the cross-section calculation  or
to reduce the output in stick-spectra type calculations (done by simply specifying a constant intensity threshold value in cm$/$molecule in the input file). The
constant intensity cut-offs are however known to
cause problems at long wavelengths,  where the density of lines is small and  each line, even weak, can
be important.  A more sophisticated method is to use the dynamic HITRAN's
intensity cut-off \citep{jt557}, defined as
\begin{equation*}
I_{\rm cut-off} = \left\{
\begin{array}{cc}
I_{\rm crit} \, (\tnu/\!\tnu_{\rm crit}) \, \tanh\left({c_2 \tnu}/\!{\,2T}\right) & {\rm
for}\ \tnu \le \tnu_{\rm crit},\\
I_{\rm crit} & \tnu > \tnu_{\rm crit},
\end{array}
\right.
\end{equation*}
where the HITRAN values for $\tnu_{\rm crit}$ and $I_{\rm crit}$ are
2000~\cm\ and $10^{-29}$ cm$/$molecule, respectively. These values are also default in \xcross\  but can be changed in the input.

\subsection{HITRAN}

\xcross\ can be used to work with the line list  in the HITRAN native format \texttt{.par}, which covers almost all its functionality. It can also be used to convert to ExoMol to HITRAN format (see Section~\ref{s:HITRAN}).

\subsection{Phoenix}

\xcross\ has the facility to output data in \textsc{Phoenix} format \citep{PHOENIX}.  In order to speed up the line-by-line calculations \textsc{Phoenix}'s atomic and molecular line lists have a compact structure, where all required properties (line positions, oscillator strengths, lower state energies and broadening parameters) are stored as 4- and 2-bytes integers. For the wavelength ($\mu$m, 4 byte-integers) this is defined as:
$$
  i_{\lambda} = {\rm int}\left(\frac{\log(\lambda)}{R_{\log}}+0.5\right),
$$
where
$$
R_{\log} = \log\left(1.0+\frac{1.0}{2000000.0}\right).
$$
The oscillator strength $gf_{\rm fi}$ for a $f\gets i$ transition, energy term values $\tilde{E}$, and broadening parameters $\gamma$ and $n$ are mapped onto 2-byte integers  according to
$$
i_{p} = {\rm int}\left(\log(p) \frac{1.0}{0.001 \,\log(10.0)}\right) + 2^{14},
$$
where $p$ is one of these properties. The integers $i_{\lambda}$, $i_{\gamma}$ and $i_n$ are then written as unformatted records with direct access, each of which containing data for 65536 lines (block-size). For molecules the broadening parameters include the reference Voigt line widths due to H$_2$ ($\gamma_{{\rm H}_{2}}$) and He  ($\gamma_{\rm He}$) and the corresponding temperature exponents $n_{{\rm H}_2}$ and $n_{\rm He}$ (see below). It should be noted that \texttt{Phoenix} uses the so-called astrophysics-convention for the nuclear statistical weights, which are related to the physics convention (adopted by ExoMol and HITRAN) as follows:
\begin{equation}\label{e:g:conv}
  g_{\rm i}^{\rm ns-astro} = \frac{g_{\rm i}^{\rm ns-phys}}{\sum_{i} g_{\rm i}^{\rm ns-phys}},
\end{equation}
where $i$ counts different nuclear statistics. For example, in case of water (H$_2$$^{16}$O), the nuclear statistics $g_{i}^{\rm ns}$ factors (physics convention) are 1 (para) and 3 (ortho), thus $g_{i}^{\rm ns}$ in the astrophysics convention are 1$/$4 (para) and 3$/$4 (ortho).
Since \texttt{Phoenix}'s partition functions are directly affected by the astrophysics convention, in order to be consistent, the ExoMol $gf_{\rm fi}$ values have to be scaled by the factor $1/4$ for water,
or $\left(\sum_{i} g_{\rm i}^{\rm ns-phys}\right)^{-1}$ in general.

\subsection{Treating non-local thermodynamic equilibrium (non-LTE)}
\label{s:non-LTE}


\xcross\ provides a simple approach to treating non-LTE environments by differentiating between the rotational and vibrational (vibronic) temperatures when calculating intensities  or partition functions (or other $T$-dependent properties).
 To this end we approximate the total energy as a sum of the vibrational (or vibronic) and rotational contributions;
\begin{equation}\label{e:Erot:Evib}
  \tE_{\varv,J,k} = \tE_{\varv}^{\rm vib} + \tE_{J,k}^{\varv,\rm rot},
\end{equation}
where $\varv$ and $k$ are generic vibrational (vibronic) and rotational quantum numbers, respectively. If the pure vibronic contributions are taken as the corresponding energy values at $J=0$ (integer spin), $J=1/2$ (non-integer spin) or the lowest $J$ allowed by the symmetry of the electronic term and the parity, corresponding to the lowest states (usually `+' or `e'). The rotational contribution is simply given by
\begin{equation}\label{e:Erot:Evib:J:k}
  \tE_{J,k}^{\varv,\rm rot} = \tE_{\varv,J,k} - \tE_{\varv}^{\rm vib}.
\end{equation}
We also assume that the rotational and vibrational modes are in corresponding (Boltzmann) LTE and that the non-LTE population of a given state (used in intensity and/or partition function calculations)  is given by
$$
F_{J,\varv,k}(T_{\rm vib},T_{\rm rot}) = e^{-c_2 \tE_{\varv}^{\rm vib}/T_{\rm vib}} e^{-c_2 \tE_{J,k}^{\varv,\rm rot}/T_{\rm rot}}.
$$
For this representation it is important to have all the vibrational and rotational quantum numbers defined in the line list, or at least for states accessed by non-LTE calculations.



\section{Line profiles}
\label{s:profile}

The line broadening is important for practical applications. While
temperature effects are commonly modelled by a Doppler line profile, pressure
broadening is more complicated. For  very high pressure regimes  Lorentzian
profiles can be used, while for  moderate pressures Voigt profiles are
generally used (see, for example, \citet{17Schreier}).

\subsection{Standard line profiles and sampling method}

The most commonly used line profiles in \xcross\ include Gaussian, Doppler, Voigt and Lorentzian.

The general Gaussian line profile is given by  \citep{13HiYuTe}
\begin{align}
\label{e:Gauss}
f^{\rm G}_{\tnu_{\rm fi}, \alpha_{\rm G}}(\tnu) = \sqrt{\frac{\ln 2}{\pi}}\frac{1}{\alpha_{\rm G}}\exp\left( -\frac{(\tnu-\tnu_{\rm fi})^2 \ln 2}{\alpha_{\rm G}^2} \right),
\end{align}
where $\tnu_{\rm fi}$  is the line centre position and $\alpha_{\rm G}$ is the Gaussian half-width at half-maximum (HWHM). The Gaussian line profile is useful to model generic spectra represented by lines with constant HWHM. The Gaussian line profile can be also used to model the microturbulence broadening by choosing $\alpha_{\rm G}$ appropriately.

The Doppler line profile  $f^{\rm D}_{\tnu_{\rm fi}, \alpha_{\rm D}}(\tnu)$ is based on the Gaussian shape defined in Eq.~\eqref{e:Gauss},
where  the Doppler HWHM is given by
\begin{align}
\label{e:alphaD}
\alpha_{\rm D} = \sqrt{\frac{2k_\mathrm{B}T\ln 2}{m}}\frac{\tnu_{\rm fi}}{c},
\end{align}
at temperature $T$ for a molecule of mass $m$.

The Lorentzian  profile is given by
\begin{equation}\label{e:lor}
 f^{\rm L}_{\tnu_{\rm fi},\gamma_L}(\tnu) = \frac{1}{\pi} \frac{\gamma_L}{(\tnu-\tnu_{\rm fi})^2+\gamma_L^2},
\end{equation}
where $\gamma_L$ is the Lorentzian line width (HWHM), given most commonly by
\begin{equation}\label{e:gamma_L}
\gamma_L = \gamma_L^0 \left(\frac{T_0}{T}\right)^{n_L} \frac{P}{P_0}.
\end{equation}
Here $T_0$ and $P_0$ are the reference temperature and pressure, respectively, $\gamma_0$ and $n_L$ are broadening parameters for a given broadener, reference HWHM and temperature exponent, respectively.

The Voigt profile is a convolution of the Doppler and Lorentazian profiles:
\begin{equation}\label{e:Voigt}
 f^{\rm V}_{\tnu_{\rm fi},\alpha_{\rm D},\gamma_{\rm L}}(\tnu)  = \frac{\gamma_{\rm L} \ln 2}{\pi^{3/2}\alpha_{\rm D}}
 \int_{-\infty}^{\infty} \frac{ e^{-y^2} \rmd y}{(\nu-y)^2+\gamma_{\rm L}^2},
\end{equation}
where we introduced a unitless variable $\nu$ given by:
\begin{equation}\label{e:nu:alphaD}
\nu = \frac{(\tnu-\tnu_{\rm fi})\ln 2}{\alpha_D}.
\end{equation}


The Lorentzian line width $\gamma_{\rm L}$ strongly depends on the molecule and is usually also state-dependent. The corresponding values must be given in the input including the specification of the broadeners and their mixing ratio. Each calculation can handle only one combination of broadeners.

Additionally, a simple box-type  line profile given by
\begin{equation}\label{e:Doppler}
 f_{\Delta \tnu }^{\rm B}(\tnu) = \left\{
 \begin{array}{cc}
   1/\Delta \tnu, & |\tnu-\tnu_{\rm fi}| \le \Delta \tnu   \\
   0, & |\tnu-\tnu_{\rm fi}| > \Delta \tnu,
 \end{array}
 \right.
\end{equation}
where $\Delta \tnu$ is the width of the box, is available.

The individual contribution from each line to the cross sections at a given frequency grid point $k$ is evaluated by sampling the corresponding line profile (see Eq.~\ref{e:cross}) a given by
$$
\sigma_{\rm fi}(\tnu_k) =  I_{\rm fi} f_{\tnu_{\rm fi}}(\tnu_k),
$$
which will be often referred to as a sampling method.  This method has the disadvantage of underestimating the opacity when too coarse grids are used which can lead to lines being  partially or completely left out. This is a typical problem for long wavelengths where the lines are narrow and far from each other, which  is usually tackled either by re-normalizing the line area, see, for example, \citet{07ShBuxx.dwarfs}, or by using a random sampling \citep{16LuMaLe}. Below we explore a different, more rigorous alternative.


In practical applications the cross sections are computed on a grid of frequencies (wavenumbers) $\tnu_i$. When the grid is not sufficiently dense, the line profiles lose their normalisation. This is usually not a problem, at least for most of the room temperature applications. However for high $T$ when billions of lines are used, this leakage can lead to a significant loss of opacity. In order to prevent this effect, \citet{13HiYuTe} suggested using an averaged intensity over a given frequency bin, where the corresponding cross section is integrated analytically. This method originally presented for the Gaussian (Doppler) line profile, is extended  here to describing Lorentzian and Voigt profiles.

\subsection{Binned Gaussian profile with analytical integrals}

An averaged (integrated) cross section over a bin $[\tnu_k-\dnu/2 \ldots \tnu_k+\dnu/2]$ from a line f$\gets$ i is given by
\begin{eqnarray}
\label{eqn:sig_ij}
\bar{\sigma}_{k}^{\rm fi} &=& \frac{I_{\rm fi}}{\dnu} \int_{\tnu_k-\dnu/2}^{\tnu_k+\dnu/2} f^\mathrm{G}_{\tnu_{\rm fi}, \alpha_{\rm fi}}(\tnu) \,\rmd \tilde{\nu}\\
&=& \frac{I_{\rm fi}}{2\dnu} \left[ {\rm erf}(x_{k,\rm if}^{+}) - {\rm erf}(x_{k,if}^-) \right],
\end{eqnarray}
where erf is the error function and
\begin{equation}
x_{k,\rm fi}^\pm = \frac{\sqrt{\ln 2}}{\alpha_{\rm D}}\left[ \tnu_k \pm \frac{\dnu}{2} - \tnu_{\rm fi} \right]
\end{equation}
are the scaled limits of the wavenumber bin centred on $\tnu_{k}$ relative to the line centre, $\tnu_{\rm fi}$, and $I_{\rm fi}$ is the line intensity in units of cm$^{-1}/$molecule~cm$^{-2}$ from Eq.~\eqref{e:int}. Here we take advantage of the fact that an analytical solution exists for the integral of the Gaussian function
\begin{equation}\label{e:erf}
  \int e^{- x^2} \rmd x = \frac{\sqrt{\pi}}{2} {\rm erf(x)} +C,
\end{equation}
where $C$ is an integration constant.
The total cross section at the frequency bin $k$ is given by a sum over all contributions from individual lines $\rm fi$:
\begin{equation}
\bar{\sigma}_k = \sum_{\rm fi} \bar\sigma_{k}^{\rm fi}
\end{equation}
and can be interpreted as an average value of the cross sections from a given frequency bin $k$. The advantage of this approach is that in definition it always gives exact integrated cross sections independent of the number of grid points used or the integration interval. Therefore it is recommended for applications where accurate integrated cross sections or absorption coefficients on coarse grids are required.
However it is known that averaged cross sections,  especially on coarse grids, can lead  to huge errors in integrated flux. Therefore for radiative transfer applications, the direct sampling methods are more accurate and should be used instead.

\subsection{Binned Lorentzian profile with analytical integrals}

Here we apply the same idea of analytical integral to the Lorentzian line profile:
\begin{eqnarray}
\label{eqn:sig_lor}
\bar{\sigma}_{k}^{\rm fi} &=& \frac{I_{\rm fi}}{\dnu} \int_{\tnu_k-\dnu/2}^{\tnu_k+\dnu/2} f^\mathrm{L}_{\tnu_{\rm fi}, \alpha_{\rm fi}}(\tnu) \,\rmd \tilde{\nu}\\
\label{eqn:sig_lor-2}
&=&  \frac{I_{\rm fi}}{\pi\dnu} \left [\arctan (y_{k,\rm fi}^+) - \arctan (y_{k,\rm fi}^-)\right],
\end{eqnarray}
where
\begin{equation}
\label{e:y_k^pm}
y_{k,\rm fi}^\pm = \frac{\tnu_k \, \tnu_{\rm fi} \, \pm \, \Delta \tnu /\! 2 }{\gamma_{\rm L}}.
\end{equation}
Here  the following integral was used:
$$
\int \frac{ \rmd x }{x^2+\gamma^2}= \frac{1}{\gamma} \arctan\left( \frac{x}{\gamma} \right) +C,
$$
where $C$ is an integration constant.
Again, the integration within each bin is done analytically which guarantees no loss of accuracy for any number of points.


\subsection{Binned Voigt profile with analytical integrals}

The two line profiles (Gaussian and Loretnzian) can be combined to produce a similar formulation for the Voigt profile, where we use the idea of Gauss-Hermit quadratures as, for example, used in \Humlicek's algorithm \citep{79Humlic}.  The Voigt convolution integral in Eq.~\eqref{e:Voigt} can be written using these quadratures as follows:
\begin{equation}
\label{e:V-quad}
 f^{\rm V}_{\tnu_{\rm fi},\alpha_D,\gamma_L}(\tnu)  = \frac{\gamma_{\rm L} \ln 2}{\pi^{3/2}\alpha_{\rm D}^2}
 \sum_{k=1}^{N_{\rm G-H}} \frac{ w_{k}^{\rm G-H}  }{(\nu-\nu_k)^2+\gamma_{\rm L}^2}.
\end{equation}
where  $\nu_k$ and $w_k^{\rm G-H}$ are the Gauss-Hermite quadrature points and weights, respectively ($k=1\ldots N_{\rm G-H} $) and $\nu$  is related to $\tnu$ via Eq.~\eqref{e:nu:alphaD}. 
In this form the computation of Voigt can be also generalised to produce the area-conserved integrals using Eq.~\eqref{eqn:sig_lor-2}:
\begin{equation}
\label{e:sigma:quad}
\sigma_{ij}^{\rm V} = \frac{I_{\rm fi}}{\pi^{3/2} \dnu} \ \sum_{k=1}^{N_{\rm G-H}} w_{k}^{\rm G-H}
\left [\arctan (y_{k,\rm fi}^+) - \arctan (y_{k,\rm fi}^-)\right].
\end{equation}
where $y_{k,\rm fi}^\pm $ is defined in Eq.~\eqref{e:y_k^pm}.
We usually take  $N_{\rm G-H} = 30$ Gauss-Hermite points. This approach does not appear to have been taken previously.

\subsection{Vectorized Voigt approximation}

\begin{figure}
\centering
\includegraphics[width=0.9\columnwidth]{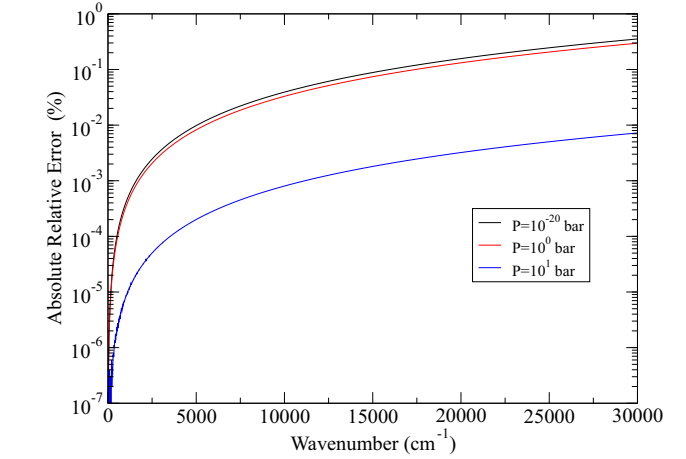}
\caption{Plot of absolute relative error at 4~\cm\ for the Voigt profile at $\tnu$ against $\tnu=1.0$ \cm\ for H$_2$O with T=5000~K and $\gamma_L$ computed from Eq.~(\ref{e:gamma_L}) with parameters $\gamma_L^0=0.0709$, $T_0=296.0$~K,~$n_L=0.5$,~$P_0=1$ and pressure $P$ at 10$^{-20}$, 10$^0$ and 10$^1$ bar. }
\label{fig:weird}
\end{figure}

Evaluation of Voigt line profile is generally one of the biggest bottlenecks in opacity calculations. Here we present a new approximate cross section algorithm for the Voigt line profile, which leads to efficient vectorization and thus fast calculations.
Our approach is based on the observation that the shape of the wings of the Voigt profile ($>4$ \cm\ from the line centre), at least for \Humlicek's algorithm, is relatively constant over the large variation of $\tnu$ as Lorentzian broadening is generally the largest contributor. For example, Figure \ref{fig:weird} shows how the wings of the Voigt profile centred at $\tnu_{\rm fi}=1$ differ from the wings of other Voigt profiles centred at all other $\tnu_{\rm fi}$ across the entire wavenuber range from 0 to 30~000~\cm\ (computed using \Humlicek's algorithm). As expected, the error grows as the Doppler HWHM (Eq. \ref{e:alphaD}) increases with transition wavenumber. However this never exceeds more than 1\% for even the lowest pressure. One of the most interesting observations is that at $10^{0}$ bar, the relative error is almost the same as the mostly Doppler profile error at 10$^{-20}$ bar. With higher pressures this error falls significantly to lower than 10$^{-2}$~\% and lower temperatures reduces this by orders of magnitudes.
It is only around the line centre, which we estimate to be within 4~\cm, that the variation of the line shape of the Voigt profile is important.

Based on this observation, the Voigt profile $f^{\rm V-V}_{\tnu_{\rm fi},\alpha_{\rm D},\gamma_{\rm L}}(\tnu)$ can be split into two parts as follows (omitting the indeces $\alpha_{\rm D},\gamma_{\rm L}$ for simplicity):
\begin{equation} \label{e:fast-v}
 f^{\rm V-V}_{\tnu_{\rm fi}}(\tnu) =
\begin{cases}
 f^{\rm V}_{\tnu_{\rm fi}}(\tnu), & |\tnu-\tnu_{\rm fi}| \leq 4\; {\rm cm}^{-1} \\
 f_{\alpha_{\rm D},\gamma_{\rm L}}^{\text{ref}}(\tnu) \,\beta_{\tnu_{\rm fi}}, & |\tnu-\tnu_{\rm fi}| > 4\ {\rm cm}^{-1}
\end{cases}
\end{equation}
where ${f_{\alpha_{\rm D},\gamma_{\rm L}}^{\text{ref}}}(\tnu)$ is a reference Voigt profile centred at $\tnu_{\rm fi}=1$~\cm:
$$
f_{\alpha_{\rm D},\gamma_{\rm L}}^{\text{ref}}(\tnu) = f^{\rm V}_{1\,{\rm cm}^{-1}}(\tnu).
$$
Here $\beta_{\tnu}$ is a parameter that is used to prevent discontinuities at $\tnu= \tnu_{\rm fi} \pm 4$~\cm\  when switching between the two profiles and is given by:
\begin{equation}\label{e:beta}
 \beta_{\tnu_{\rm fi}} =
 \frac{f^{\rm V}_{\tnu_{\rm fi}}(4\,{\rm cm}^{-1})}{f_{\alpha_{\rm D},\gamma_{\rm L}}^{\text{ref}}(4\,{\rm cm}^{-1})}.
\end{equation}
This parameter is included for completeness and is generally set to $\beta=1$ for a performance boost as the discontinuities are not visible at most scales for a single transition and invisible once the whole spectrum is considered.
For a given set of pressure broadening parameters  $\gamma_{\rm L}$ we pre-compute
a set of points defining the wings $f_{\alpha_{\rm D},\gamma_{\rm L}}^{\text{ref}}$ and then simply select a relevant set. Therefore the only palce where real Voigt calculation needs to be done is around the centre. Additionally, (if used) $\beta_{\tnu_{\rm fi}}$ needs to be calculated at the boundary, which completes the evaluation of the given profile.

\begin{figure}
\centering
\includegraphics[width=0.9\columnwidth]{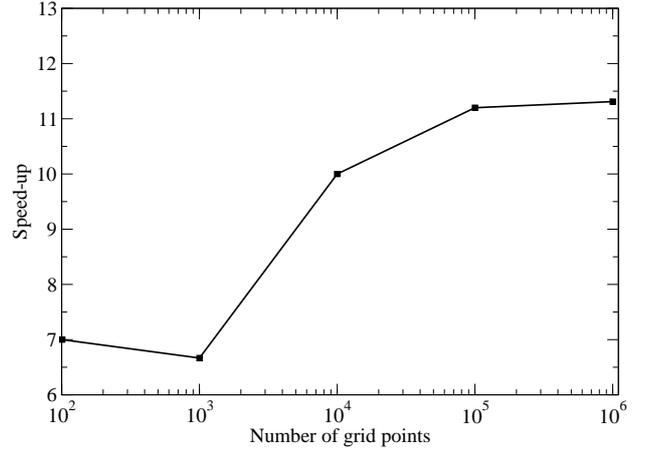}
\caption{Speed up, Eq.~(\ref{e:speedup}), for the Vectorized Voigt method against the standard Voigt  (\Humlicek) computed on varying wavenumber grid sizes ($N$) using the BT2 \citep{jt378} water line list computed at T=1900~K, P=1 bar and wavenumber range between 0 \cm\ and 300 \cm\ }
\label{fig:vecvoigt-speedup}
\end{figure}

The algorithm is based on the  \citep{79Humlic} approximation for  the Voigt profile in Eq.~\eqref{e:Voigt}, which  is the main method used by \xcross. The \Humlicek\ algorithm is called only for the regions within 4 \cm\ from the line centre. Using the conventionally used Lorentz cutoff of 25 \cm, this means that only up to 8\% of the calculation is computationally demanding giving a
theoretical speed up of 12.5 times. This is illustrated in Figure \ref{fig:vecvoigt-speedup}, which shows speed up using our Vectorized Voigt algorithm when applied to the region of 0.0--300 \cm\  of the BT2 water linelist \citep{jt378} at T=1900~K and P=1 bar. The speed up $S$ for $N$ points used to bin the wavenumber grid is defined as:
\begin{equation}\label{e:speedup}
 S^{N} = \frac{T^{N}_0}{T^{N}_{\rm V-V}},
\end{equation}
where $T^{N}_0$ is the time required for a standard \Humlicek\ computation on a wavenumber grid $N$ and $T^{N}_{\rm V-V}$ is the time required using the Vectorized Voigt method. The speed up converges to a maximum value of about 11 times compared to the standard \Humlicek\ calculation, close to the predicted maximum speed up.

This procedure is also efficiently vectorized. Firstly, for the inner part (top of Eq. (\ref{e:fast-v})), which is symmetric,  only one half is computed. The other
half is then merely looped through backwards and applied to the grid, requiring
only to multiply by the absorption coefficient (emissivity) and to add to the opacity grid. The second vectorization occurs when dealing with the second part of Eq.
\eqref{e:fast-v}. Here again, only a multiplication  by the intensity and add to the opacity grid is required. These two loops are vectorized through the Fused-Multiply-Add (FMA) instruction.

\begin{figure}
\centering
\includegraphics[width=0.9\columnwidth]{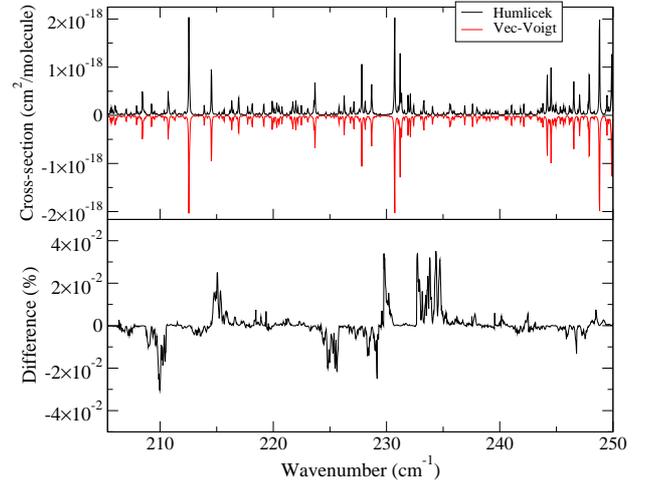}
\caption{Top plot: Comparison of cross section calculations for the BT2 water line list \citep{jt378} between the standard \Humlicek\ and
the Vectorized Voigt method with $T=$1900~K. Bottom plot: Percentage difference between the \Humlicek\ and Vectorized Voigt method. The calculations used no intensity threshold and a wavenumber bin of 0.1 \cm. }
\label{fig:vecvoigt-cross-section}
\end{figure}

Figure \ref{fig:vecvoigt-cross-section} presents an illustration, where both the Vectorized and standard (\Humlicek) Voigt methods  were used to generate cross sections of water from the BT2 line list for T=1900~K and P=1 bar. The new algorithm captures all features with the total opacity for the range shown differing by only by 10$^{-6}$ cm$^{2}$ molecule$^{-1}$.

Lastly, for a full opacity calculation between 0.0--30~000~\cm, Table \ref{tab:total_vec_time} shows that using no intensity threshold with the Vectorized Voigt method is almost 3.5 times faster than the full \Humlicek\ method at 10$^{-30}$ cm molecule$^{-1}$ thresholding. Comparing like for like, the Vectorized Voigt is around 10 to 12 times faster compared to the standard \Humlicek\ method.

\begin{table}
\centering
\caption{Time taken (s) for differing methods and intensity thresholds (cm molecule$^{-1}$)
to compute opacities using the 500 million transitions of the BT2 water line list \citep{jt378} between 0 and 30~000 \cm\
with a wavenumber binning of 0.1 \cm. I/O time is not considered. Test was performed on single core on an Intel Core i5-2500 CPU at 3.30GHz with 16 GB RAM}
\label{tab:total_vec_time}
\begin{tabular}{lcr}
\hline\hline				\\
Method    & Threshold  & Time(s)   \\
\hline				\\
Vec-Voigt & 0         & 251.2     \\
\Humlicek\  & 0         & 2775.6    \\
Vec-Voigt & 10$^{-30}$  & 70.0       \\
\Humlicek\  & 10$^{-30}$  & 832.6     \\
\hline
\end{tabular}
\end{table}

Future development of the algorithm will look into automatically tuning the distance from the line centre depending on the temperature and pressure parameters given.

\subsection{Binned Vectorized Voigt with the line area preserved}

Considering the importance of preserving integrated cross section in many applications, we also provide an alternative version of the Vectorized Voigt, based on re-normalization of the line area. During the precomputation stage of the Vectorized Voigt method, the total sum for all points ($\Sigma_{\alpha_{\rm D},\gamma_{\rm L}}$) that lie above  $|\tnu-\tnu_{\rm fi}| > 4$~\cm\ is computed and stored alongside the reference Voigt profile. When computing the Vectorized Voigt on a transition, the central \Humlicek\ region is evaluated into a temporary array and its sum is added to $\Sigma_{\alpha_{\rm D},\gamma_{\rm L}}$. After which the scaled absolute intensity $\tilde{I}_{fi}$ is computed as:
\begin{equation}
 \tilde{I}_{fi} = \frac{I_{fi}}{\Sigma_{\alpha_{\rm D},\gamma_{\rm L}}}
\end{equation}
Both the temporary \Humlicek\ array and reference Voigt is applied to the opacity grid with the scaled intensity $\tilde{I}_{fi}$.
\begin{table}[]
\centering
\caption{Table showing percentage relative error between the total summed absolute intensity and the total integrated intensity for BT2 water line list computed between 0 and 300 \cm\ at $T$=1900~K and $P$=1 bar at various wavenumber binnings. The total integrated intensities are computed using \Humlicek\ (H), Vectorized Voigt (VV) and the Normalized Vectorized Voigt (VVN).}
\label{tab:re-err}
\begin{tabular}{crrr}
\hline\hline
& & Error & \\
Bin (\cm)  & H(\%) & VV (\%) & VVN (\%)   \\
\hline \\
10.00   & 41.62    & 41.63               & 0.01                   \\
1.00    & 37.59    & 37.59               & 0.73                   \\
0.10  & 1.66     & 1.66                & 0.17                   \\
0.01 & 0.07     & 0.07                & 0.01                   \\
\hline
\end{tabular}
\end{table}
Whilst not a proper treatment of area conservation as that given by Eq. (\ref{e:sigma:quad}), it serves as a reasonable approximation and, as shown in Table \ref{tab:re-err}, gives good results within 1\% of the total summed absolute intensity for even large wavenumber bins. To our knowledge, this method does not appear to be reported before.


\subsection{Broadening parameters}

The Voigt profile as a convolution of Doppler and Lorentzian profiles requires
definition of the corresponding line widths (HWHM), $\alpha_{\rm D}$ (see Eq.~\eqref{e:alphaD}) and
$\gamma_{\rm L}$, given by Eq.~\eqref{e:gamma_L}.
The Doppler parameter  $\alpha_{\rm D}(T)$ is easy to deal with. It does
not depend on the molecular states, only the line position and can be always
computed on the fly. The Lorentzian (Voigt) parameters $
\gamma_0(P_0,T_0)$ and  $n_L$ however are very different for different molecules. Besides
they show a pronounced dependence on the state quantum numbers, with the
rotational ($J$) state dependence being  the strongest.

The two-file format of the ExoMol database requires special structure for the
broadening parameters. Instead of using the conventional line-by-line approach employed by
spectroscopic databases such in HITRAN \citep{jt691}
or GEISA \citep{jt636}, where the pressure  broadening is specified for the each transitions,
ExoMol's broadening parameters are stored in separate files with the extension
\texttt{.broad} \citep{jt631}. This structure is justified for most applications as
the same parameters are usually used for a large number of different
transitions. The latter is either due to the absence of broadening information
on all the lines or due to the weak dependence of these parameters for different
states. This structure was recently implemented for a number of molecules including H$_2$O,
CH$_4$ and HCN \citep{17BaHiYu,jt662}.
Table~\ref{tab:broad_CS_example} shows an extract from the \texttt{.broad} file for CS as
an example. Each line in \texttt{.broad} has the following structure: type (\texttt{a0}, \texttt{a1},
$\ldots$),  $ \gamma_0(P_0,T_0)$,  $n_L$ and quantum numbers defined by the type.

Currently \xcross\ supports three following broadening schemes, constant, \texttt{a0} and \texttt{a1}, depending on the rigorous quantum numbers $J'$ and $J''$. The simplest case
is when $\gamma_0(P_0,T_0)$ and $n_L$ are constant and the \texttt{.broad} data
is not required. The \texttt{a0} type corresponds to the $J$-dependence only. In this case the
4th column in the \texttt{.broad} file contains the $J$ values. The $J$ quantum
number is a mandatory quantity in the ExoMol format (column 4 in \texttt{.states}) and is
therefore relatively straightforward to handle. A similar scenario (\texttt{a1}) is when
the broadening depends on the upper $J'$ (column 5 in \texttt{.broad}) and lower
$J''$ (column 4) rotational quantum numbers. All other broadening schemes
involve dependence on some non-rigorous quantum numbers (\lq labels'), such as vibrational $\varv$
 or rotational $K$. The non-rigorous quantum numbers and their position in the
\texttt{.states} file are molecule dependent and thus need to be specified. This
information can be found in the ExoMol's \texttt{.def} (API) file. The current version of
\xcross\ supports rigorous quantum numbers only and therefore does not
require interfacing with the ExoMol database.

\begin{table*}
\caption{Air \texttt{.broad} file for
$^{12}$C$^{32}$S: portion of the file (upper part); field specification (lower part).}
\label{tab:broad_CS_example} \footnotesize
\begin{center}
\begin{tabular}{llll}
\hline
a0 & 0.0860 & 0.096 & 0 \\
a0 & 0.0850 & 0.093 & 1 \\
a0 & 0.0840 & 0.091 & 2 \\
a0 & 0.0840 & 0.089 & 3 \\
a0 & 0.0830 & 0.087 & 4 \\
... & & & \\
a0 & 0.0720 & 0.067 & 35 \\
a0 & 0.0720 & 0.066 & 36 \\
... & & & \\
\end{tabular}
\begin{tabular}{llll}
\hline
Field & Fortran Format & C format & Description \\
\hline
code & A2 & $\%$2s & Code identifying quantum number set following $J''$* \\
$\gamma_0(P_0,T_0)$ & F6.4 & \%6.4f & Lorentzian half-width at reference
temperature and pressure in \cm/bar \\
$n_L$ & F5.3 & \%5.3f & Temperature exponent \\
$J''$ & I7/F7.1 & $\%$7d & Lower $J$-quantum number \\
\hline
\end{tabular}
\end{center}
\noindent
*Code definition:
a0 = none
\end{table*}

\subsection{Mixtures of broadeners}

We consider different broadeners to be independent and their effect additive. Thus the total
value of $\gamma_{\rm L}$ is  a weighed sum of $\gamma_{i}^{\rm L}$ from
each broadener as given by:
$$
\gamma_{\rm L} = \sum_{i} \gamma_{i}^{\rm L} \rho_i
$$
where $\rho_i$ is the fraction portion of the $i$th broadener. Here we used the
fact that the cross sections from each lines are additive and thus the line
profile can be represented as a weighted average of lines broadened by
different species.

\subsection{Off-set}

Even though, at least in principle, a line profile has infinite spread,  in practical calculations a frequency (or wavelength) cut-offs must be applied to limit the calculation region to around the line centre only. Not only does this influence the computation time and the accuracy of
cross sections, but it is also assumed in some applications as a point of convention. For
example, water cross section are conventionally taken to have a
25~\cm\ cutoff, with far-wing contributions outside this region assumed
to form part of the so-called water-continuum \citep{12ShPtRa}. 25~\cm\ is the default cut-off value in \xcross, alternatively it is specified in the input file.

\subsection{Super-lines}

The super-line approach is an efficient method for describing a molecular broadened
continuum originally proposed by \citet{TheoReTS} and was recently studied in detail by \citet{jt698}.
The super-lines are constructed as temperature-dependent intensity
histograms as follows
(see also detailed discussion by \citet{TheoReTS}).  We divide the wavenumber
range $[\tilde\nu_{A},\tilde\nu_{B}]$  into $N$ frequency bins, each centred
around a grid point $\tilde\nu_k$. For each $\tilde\nu_k$ the total absorption
intensity ${I}_k(T)$ is computed as a sum of absorption line intensities
$I_{\rm fi}$, as in Eq.~\eqref{e:int}, from all f~$\to$~i transitions   falling into the
wavenumber bin $[\tilde\nu_k-\Delta \tilde{\nu}_k/2 \ldots \tilde\nu_k+\Delta
\tilde{\nu}_k/2 ] $ at the given temperature $T$.
Each grid point $\tilde\nu_k$ forms a super-line of an artificial transition
with an effective absorption intensity ${I}_k(T)$. The super-line lists are
given in a two-column format $\{\tilde\nu_k, {I}_k(T)\}$ with pre-computed
intensities ${I}_{k}$, in the same format as used to store ExoMol cross-sections
\citep{jt631}. The filename have the extension \texttt{.super}.  The super-lines
approach does not require that histograms are of the same widths $\Delta
\tilde{\nu}_k$ and can accept non-equidistant grids as well, see below.

The histograms in \xcross\ can be produced as cross sections using the
\verb!Bin!-option in the input file (see Manual), which is basically just a sum
of all intensities within a given bin $i$.
Ones the histograms are computed (in the standard cross section two-column
format), they can be treated as normal line lists. In this case the .states file is not
needed as all the information has been already included into the line position and intensity.
Moreover, since the states-specific information is completely lost from the line characteristics,
the state-dependent line profiles can  not be used for temperature/pressure broadening. Doppler line profiles require no information on the upper/lower states and are not restricted.
However for the Voigt pressure broadening parameters, which usually depend (at least) on $J$, only  constant values of $\gamma_0$ and $n_L$ (see Eq.~\eqref{e:gamma_L}) can be used in conjunction with  super-lines. For this reason the super-lines are recommended for description of  featureless continuum produced from the weaker lines only. The stronger lines should be treated as usual, line-by-line.

\subsection{User-defined profiles}

New line profiles, see \citet{jt584} for example, can be easily implemented to \xcross\ by the user. A detailed description is provided in the manual. The HITRAN option in \xcross\ can be used as an example.

\section{Calculation protocol}
\label{s:protocal}

\begin{figure}
\center
\includegraphics[width=0.94\columnwidth]{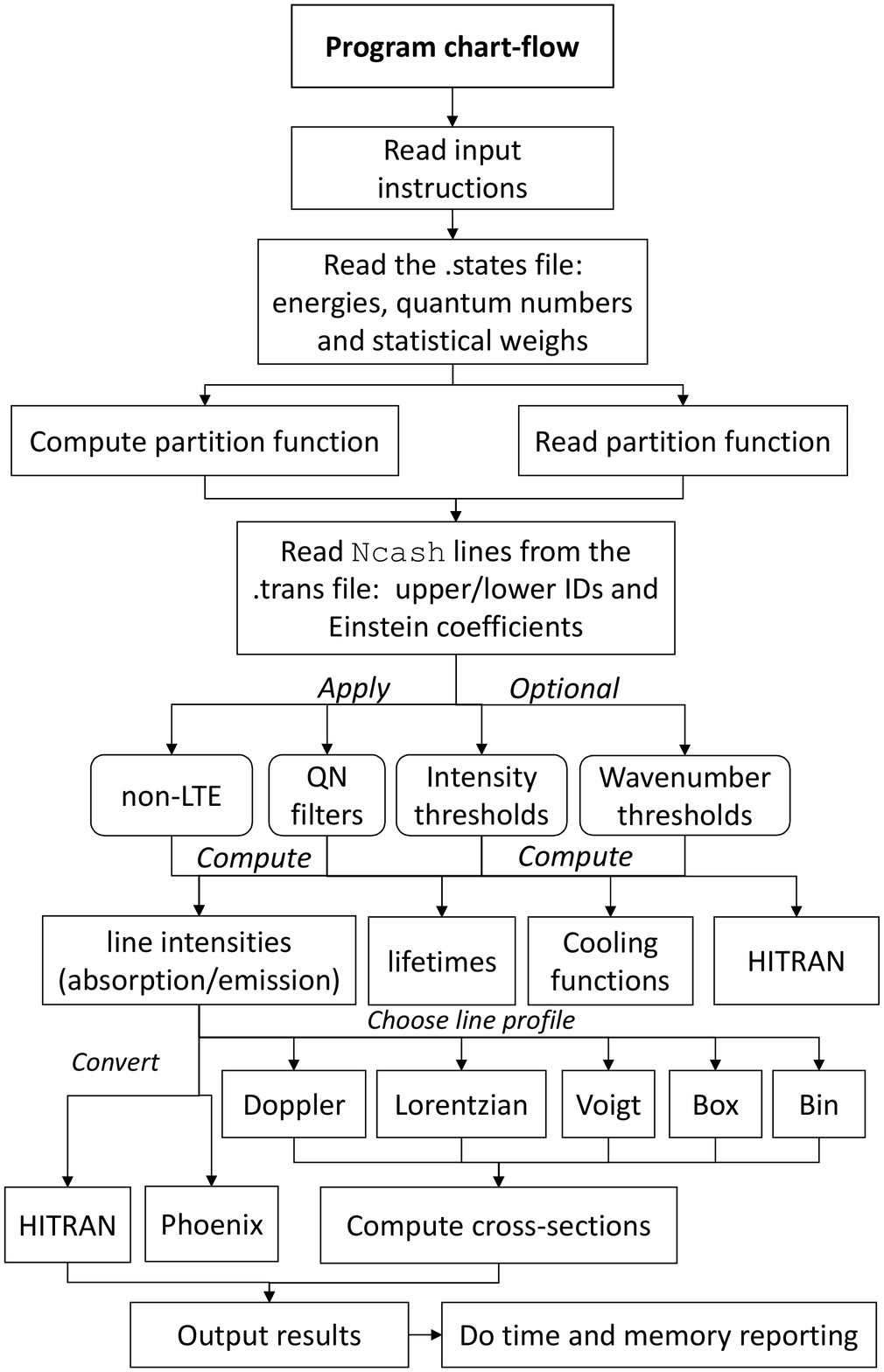}
\vspace{1cm}
\caption{\xcross\ program work-flow.}
\label{f:chart}
\end{figure}

The typical \xcross\ calculation includes the following steps (see Fig.~\ref{f:chart}):
\begin{enumerate}
  \item Read input instruction;
  \item Read the .states file: energies, quantum numbers and statistical weighs;
  \item Compute the partition function (if required);
  \item Read $N$ lines with upper/lower IDs and the Einstein coefficient  lines from the .trans file;
  \item Apply filters;
  \item Compute line intensities (absorption coefficients or emissivities, if required).
  \item Compute cross sections on a grid of wavenumbers (if required);
  \item Compute lifetimes (if required);
  \item Compute cooling functions (if required);
  \item Print the cross sections (stick spectra, life times, cooling functions) into a separate file;
  \item Do time and memory reporting.

\end{enumerate}

\section{Data formats}
\label{s:format}

\xcross\ currently takes in input in either ExoMol or HITRAN format. It can provide output in these formats and in the format used by the Phoenix radiative transport code \citep{PHOENIX}. These formats are discussed in turn below.

\subsection{ExoMol format}

A line list is defined as a catalogue of transition frequencies and
intensities \citep{12TeYuxx.db}. In the basic ExoMol format
\citep{13HiYuTe}, adopted by \xcross, a line list has a compact
structure consisting of two files: `States' and `Transitions'; an
example for the list NOname line list for $^{14}$N$^{16}$O
\citep{jt686} is given in Tables~\ref{table:states} and
\ref{tab:trans}. The `States' (\texttt{.states}) file contains energy
term values supplemented by the running number $n$, total degeneracy
$g_n^{\rm tot}$, rotational quantum number $J_n$ (all obligatory
fields), other quantum numbers and labels (both rigorous and not
rigorous), lifetimes and Land\'{e} $g$-factors. For example for a
generic open-shell diatomic molecule, the quantum numbers include
$\upsilon$, $\Lambda$, parity ($\pm$), $\Sigma$, $\Omega$ and the
electronic state label (e.g.  \verb!X2Sigma+!) \citep{Duo}. The
`Transitions' (\texttt{.trans}) file contains three obligatory
columns, the upper and lower state indexes $n_f$ and $n_i$ which are
running numbers from the `State' file, and the Einstein coefficient
$A_{\rm fi}$. For the convenience it also sometimes provides the
wavenumbers $\tilde{\nu}_{fi}$ as the column 4. The line list in the
ExoMol format can be used to simulate absorption or emission spectra
for any temperature in a general way.


\begin{table*}
\caption{Extract from the states file of the $^{14}$N$^{16}$O line list.   }
\label{table:states}
\tt
\footnotesize
 \begin{tabular}{crccccccrrrrrc}
 \hline
$n$ & Energy (\cm)	&	$g_n$	&	$J_n$ & $\tau$ & $g_J$	&	+/-	&	e/f	&	State	&	$\varv$	&	${\Lambda}$	&	${\Sigma}$	&	${\Omega}$ 	\\
\hline\hline
       1  &        0.000000  &    6  &     0.5  &   inf       &   -0.000767  &     +    &   e    &     X1/2    &      0  &      1  &    -0.5  &     0.5    \\
       2  &     1876.076228  &    6  &     0.5  &   8.31E-02  &   -0.000767  &     +    &   e    &     X1/2    &      1  &      1  &    -0.5  &     0.5    \\
       3  &     3724.066346  &    6  &     0.5  &   4.25E-02  &   -0.000767  &     +    &   e    &     X1/2    &      2  &      1  &    -0.5  &     0.5    \\
       4  &     5544.020643  &    6  &     0.5  &   2.89E-02  &   -0.000767  &     +    &   e    &     X1/2    &      3  &      1  &    -0.5  &     0.5    \\
       5  &     7335.982597  &    6  &     0.5  &   2.22E-02  &   -0.000767  &     +    &   e    &     X1/2    &      4  &      1  &    -0.5  &     0.5    \\
       6  &     9099.987046  &    6  &     0.5  &   1.81E-02  &   -0.000767  &     +    &   e    &     X1/2    &      5  &      1  &    -0.5  &     0.5    \\
       7  &    10836.058173  &    6  &     0.5  &   1.54E-02  &   -0.000767  &     +    &   e    &     X1/2    &      6  &      1  &    -0.5  &     0.5    \\
       8  &    12544.207270  &    6  &     0.5  &   1.35E-02  &   -0.000767  &     +    &   e    &     X1/2    &      7  &      1  &    -0.5  &     0.5    \\
       9  &    14224.430238  &    6  &     0.5  &   1.21E-02  &   -0.000767  &     +    &   e    &     X1/2    &      8  &      1  &    -0.5  &     0.5    \\
      10  &    15876.704811  &    6  &     0.5  &   1.10E-02  &   -0.000767  &     +    &   e    &     X1/2    &      9  &      1  &    -0.5  &     0.5    \\
      11  &    17500.987446  &    6  &     0.5  &   1.01E-02  &   -0.000767  &     +    &   e    &     X1/2    &     10  &      1  &    -0.5  &     0.5    \\
      12  &    19097.209871  &    6  &     0.5  &   9.41E-03  &   -0.000767  &     +    &   e    &     X1/2    &     11  &      1  &    -0.5  &     0.5    \\
      13  &    20665.275246  &    6  &     0.5  &   8.83E-03  &   -0.000767  &     +    &   e    &     X1/2    &     12  &      1  &    -0.5  &     0.5    \\
      14  &    22205.053904  &    6  &     0.5  &   8.35E-03  &   -0.000767  &     +    &   e    &     X1/2    &     13  &      1  &    -0.5  &     0.5    \\
      15  &    23716.378643  &    6  &     0.5  &   7.94E-03  &   -0.000767  &     +    &   e    &     X1/2    &     14  &      1  &    -0.5  &     0.5    \\
      16  &    25199.039545  &    6  &     0.5  &   7.59E-03  &   -0.000767  &     +    &   e    &     X1/2    &     15  &      1  &    -0.5  &     0.5    \\
\hline

\end{tabular}
\mbox{}\\
{\flushleft
$n$:   State counting number.     \\
$\tilde{E}$: State energy in \cm. \\
$g_n$:  Total statistical weight, equal to ${g_{\rm ns}(2J_n + 1)}$.     \\
$J_n$: Total angular momentum.\\
$\tau$: Lifetimes (s$^{-1}$).\\
$g_J$: Land\'{e} $g$-factor \\
$+/-$:   Total parity. \\
$e/f$:   Rotationless parity. \\
State: Electronic state.\\
$\varv$:   State vibrational quantum number. \\
$\Lambda$:  Projection of the electronic angular momentum. \\
$\Sigma$:   Projection of the electronic spin. \\
$\Omega$:   $\Omega=\Lambda+\Sigma$, projection of the total angular momentum.\\

}

\end{table*}

\begin{table}
\caption{Extract from the transitions file of the $^{14}$N$^{16}$O  line list. }
\label{tab:trans}
  \centering
  \tt
  \footnotesize
\begin{tabular}{rrrr}
\hline
\multicolumn{1}{c}{$f$}	&	\multicolumn{1}{c}{$i$}	& \multicolumn{1}{c}{$A_{\rm fi}$ (s$^{-1}$)}	&\multicolumn{1}{c}{$\tilde{\nu}_{\rm fi}$ (\cm)}	\\
\hline\hline
       14123   &    13911&  1.5571E-02    &    10159.167959 \\
       13337   &    13249&  5.9470E-06    &    10159.170833 \\
        1483   &     1366&  3.7119E-03    &    10159.177466 \\
        9072   &     8970&  1.1716E-04    &    10159.177993 \\
        1380   &     1469&  3.7119E-03    &    10159.178293 \\
       14057   &    13977&  1.5571E-02    &    10159.179386 \\
       10432   &    10498&  4.5779E-07    &    10159.187818 \\
       12465   &    12523&  5.4828E-03    &    10159.216008 \\
       20269   &    20286&  1.2448E-10    &    10159.227463 \\
       12393   &    12595&  5.4828E-03    &    10159.231009 \\
        2033   &     2111&  6.4408E-04    &    10159.266541 \\
       17073   &    17216&  4.0630E-03    &    10159.283484 \\
        5808   &     6085&  3.0844E-02    &    10159.298459 \\
        5905   &     5988&  3.0844E-02    &    10159.302195 \\
       13926   &    13845&  1.5597E-02    &    10159.312986 \\
\hline
\end{tabular}

\noindent
 $f$: Upper  state counting number;\\
$i$:  Lower  state counting number; \\
$A_{\rm fi}$:  Einstein-A coefficient in s$^{-1}$; \\
$\tilde{\nu}_{\rm fi}$: transition wavenumber in \cm.\\

\end{table}

\subsection{HITRAN}
\label{s:HITRAN}

 The current ``HITRAN format'' is fully specified
in Table 1 of the 2004 edition of HITRAN \citep{jt350}. This format, which is also used
for the current release of the related high-temperature database HITEMP \citep{jt480}, has been implemented here.

Although the HITRAN format is widely adopted as a {\it de facto} standard, we advise some caution
before adopting it. The format is rather verbose and can become extremely unwieldy as a means of
representing large line lists. The format is highly tuned towards Earth atmosphere application (e.g. in its choice pressure broadening parameters and temperature ranges) and is therefore rather inflexible for other applications. HITRAN themselves have recognised these issues and have introduced their own web-based interface HAPI \citet{HAPI} to act as front end and to perform data compression. The database itself has moved to an online-version which provides much more flexibility than the 2004 format
\citet{jt559}.

\subsection{Improving data processing}

Both the cross-section and intensity steps (see Fig.~\ref{f:chart})  are OpenMP
parallelized.
Users can specify the number of processors requested, which is otherwise set to 1 (no parallelization). In order to make reading and processing data from the \texttt{.trans} file more efficient,
\xcross\ reads line transitions in chunks of $N$  lines, not line-by-line.
`Caching' these records into RAM allows for the parallelization for both the
transition filtering and of the computation of line-profiles. Each thread is
given their own version of the opacity grid to perform work independently
without the usage of atomic operations or mutex locks. The total opacity grid
can be retrieved at the end of the program run combining all threads' opacity
arrays.  This number $N$ is either specified in the input file or estimated based on the memory
available on the system (default). The number of processors must be specified in the input as well (see below on the memory handling).

\subsection{Filters}

\begin{figure}
\begin{center}
\includegraphics[angle=0, width=0.95\columnwidth]{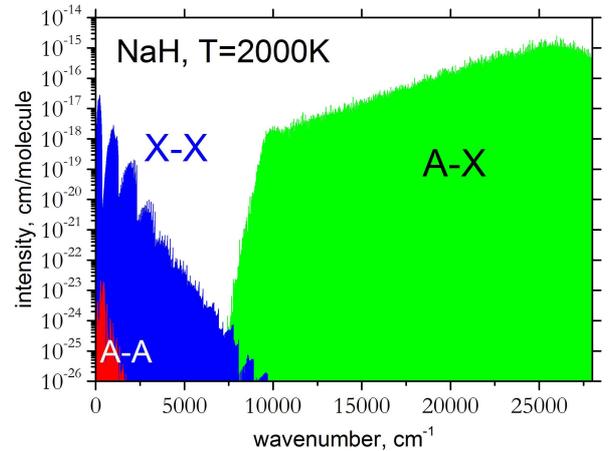}
\caption{Overview of the absorption line intensities of NaH at $T$=2000~K computed using the line list of \protect\citet{15RiLoYu.NaH}. \label{f:NaH}}
\end{center}
\end{figure}

\xcross\ allows the selection of specific bands/states when computing intensities using the `filter' option. The filters are based on the column-numbers containing the corresponding quantum labels of the upper and lower states. For example, the vibrational quantum number $\varv$ in the  NOname line list  is given in the column number 10 (see Table~\ref{table:states}), which can be used  to generate absorption cross section of NO for the overtone band $\varv=5$, i.e. for transitions  between $\varv'=5$ and $\varv''=0$ of NO, by referring in the input the corresponding values from  the column 10 (see Manual for details).
Another typical example is to generate cross sections for specific electronic bands, see Fig.~\ref{f:NaH}, where an overview of three absorption electronic bands X--X, A--X, A--A of NaH is shown \citep{15RiLoYu.NaH}.

The filter-feature will work even if not all states are assigned. According to the ExoMol convention, the string \texttt{NaN} (with any combination of upper an lower cases) is used for missing quantum labels. Thus `NaN' in this case will be effectively used by \xcross' filter as a quantum label.

\subsection{Units}

The default units of \xcross\ are listed in Table~\ref{t:units}.  Microns ($\mu$m) can be optionally used for wavelength as alternative to wavenumber (default). Pressure does not have designated units; it is assumed to have the same units as of the parameter $P_0$ defining the broadening parameter $\gamma$, see Eq.~\eqref{e:gamma_L}.

\begin{table}
\caption{Units used by \xcross\ }
\label{t:units}
\begin{tabular}{ccccc}
  \hline
  \hline
  Quantity & Units \\
  \hline
Wavenumber & \cm \\
Wavelength & $\mu$m \\
Temperature & K \\
Pressure & units of $P_0$ \\
Absorption coefficient & cm$/$molecule \\
Absorption cross sections & cm$/$[molecule cm$^{-1}$] \\
Emissivity &  erg$/$s~sr~molecule  \\
Specific heat & $J K^{-1}$ mol$^{-1}$ \\
  \hline
  \hline
\end{tabular}
\end{table}

\subsection{Memory handling}

The program records and controls the memory used at all processors. For proper
control, the user is requested to specify the memory available on the machine
in Gb or Mb. This number is used, for example, to estimate the
number of transition lines from \texttt{.trans} processed simultaneously. At
the end of the program a memory usage report is given.

\section{Program repository}

The \xcross\ code together with manual and input examples are freely available from the ExoMol website \footnote{\url{www.exomol.com}}, CCPForge
\footnote{\url{https://ccpforge.cse.rl.ac.uk/gf/project/exocross/}} or GitHub
\footnote{\url{https://github.com/Trovemaster/exocross}}.

\section{Conclusion}
\label{s:conclusion}

We present a new Fortran program \xcross\ to compute different spectroscopic properties of molecules using spectral line lists. The program has being actively used by ExoMol to generate absorption cross sections using the ExoMol line lists available at \url{www.exomol.com}. In order to work  with huge sizes of some line lists, \xcross\ is optimized for efficient usage of parallelizm and vectorisation. Our new Voigt algorithm (Vectorized Voigt) is designed to be fast and accurate.

The program can  easily be extended by users with their profiles or other functionality.

We are planning to provide production of $k$-coefficients as part of \xcross\ in the future; integrate the API  via the ExoMol \texttt{.def} file; reading the partition function from an ExoMol \texttt{.pf} file; implement a non-LTE model, which does not require definition of non-rigorous quantum numbers (see Section~\ref{s:non-LTE}).


\section*{Acknowledgements}

We would like to acknowledge help of Derek Homeier and France Allard with  Phoenix format. We thank Ingo Waldmann and Marco Rocchetto for testing \xcross.
\xcross\ uses the Fortran~90 input parsing module input.f90 supplied by Anthony J. Stone, which is gratefully acknowledged. This work was supported by the UK Science and Technology Research Council (STFC) No. ST/M001334/1 and the COST action MOLIM No. CM1405.  This work made extensive use of UCL's Legion and STFC's DiRAC@Darwin high performance computing facilities.

\bibliographystyle{aa}

\end{document}